\documentclass[a4paper,11pt]{article}
\pdfoutput=1
\usepackage{jheppub}
\usepackage{hyperref}
\usepackage[T1]{fontenc}
\usepackage{booktabs,xcolor,enumitem,tikz}
\usetikzlibrary{arrows.meta,calc,positioning}
\usepackage{hyperref}
\hypersetup{colorlinks=true,linkcolor=blue!50!black,citecolor=blue!50!black,urlcolor=blue!50!black}

\title{\boldmath JetCoRD: Reliability-Gated Correction of Jet Taggers over Two Frozen Teachers}

\author[a,b]{Liu-Long Gao,}
\author[a,b]{Zheng-Kun Huang,}
\author[a,b]{Xiao-Wei Jiang,}
\author[a,b]{Jia-Feng Li}
\author[a,1]{and Gong-Xing Sun\note{Corresponding author.}}

\affiliation[a]{Institute of High Energy Physics, Chinese Academy of Sciences,\\Beijing 100049, China}
\affiliation[b]{University of Chinese Academy of Sciences,\\Beijing 100049, China}

\emailAdd{gaoll@ihep.ac.cn}
\emailAdd{huangzk@ihep.ac.cn}
\emailAdd{jiangxw@ihep.ac.cn}
\emailAdd{jfli@ihep.ac.cn}
\emailAdd{sungx@ihep.ac.cn}

\abstract{Knowledge distillation for jet taggers is usually cast as compression toward a near-oracle teacher. We instead train a small reliability-gated correction layer jointly over two frozen teachers --- a GN2-type ATLAS flavour tagger (3 classes) and CMS ParT (10 classes) --- reading only their frozen penultimate embeddings and logits, so it adds negligible marginal cost. A single per-sample reliability signal reweights the distillation loss, drives a prototype-based repair of the teachers' errors, and anchors the inference-time gate that mixes teacher and student logits. Zero-initialised, the 82\,k-parameter student starts as a bit-exact copy of the teacher --- a provable invariant on the training distribution --- and departs only where the data support it. Under a statistically-grounded evaluation, with binomial, Poisson and paired uncertainties on every table, it reproduces both teachers at $\geq\!99\%$ agreement and matches their overall accuracy while redistributing it toward their weakest classes: the JetClass per-class accuracy of $H\!\to\!4q$, QCD and $H\!\to\!bb$ rises by $+0.8$, $+0.4$ and $+0.2$ percentage points ($|z|>7$), offset by comparable drops elsewhere. Oracle heads on the frozen embedding recover a budget consistent with zero in both domains, and a conformal-risk-control analysis certifies the false-accept rate. This full re-evaluation supersedes an earlier, incorrectly-evaluated set of numbers; code and models are available at \url{https://github.com/sysu17363020/JetCoRD}.}

\begin{document}
\maketitle
\flushbottom

\section{Introduction}
Jet tagging, which classifies collimated hadronic showers originating from underlying partons in high-energy collisions, serves as a foundational tool for nearly all physics measurements and new physics searches at the Large Hadron Collider (LHC). The state-of-the-art flavor tagger \textbf{GN2} ~\cite{GN2} at the ATLAS experiment achieves superior performance via a graph neural network architecture operating on charged particle constituents. Meanwhile, the general-purpose multi-class tagger \textbf{ParT} ~\cite{ParT}, trained on the public CMS \textbf{JetClass} dataset, reaches sub-percent class-conditional classification errors across ten physics processes. Despite their remarkable performance, both technical approaches share three key practical limitations:
\begin{enumerate}
\item \textbf{Prohibitive experiment-specific training costs.} The GN2 and ParT comprise millions of parameters and require extensive GPU training iterations on detector-specific experimental data, so every new detector, label definition, or calibration forces a fresh and expensive training campaign.
\item \textbf{No shared component across detectors.} A tagger trained on ATLAS data does not generalize to CMS data, and vice versa. Today each collaboration independently develops and maintains its own suite of jet tagging models, and there is no shared, lightweight component that can sit on top of two different teachers and serve both.
\item \textbf{Teacher errors are propagated by distillation.} The usual way to reduce model size --- knowledge distillation ~\cite{Hinton} --- assumes the teacher model delivers near-perfect, ground-truth-level classification outputs. A realistic ATLAS-type flavor tagger instead errs on roughly a quarter of jets (our teacher reaches $73.9\%$ overall accuracy on its own flavor-simulation test set), and conventional KL-distillation pulls the student toward whatever errors the teacher makes.
\end{enumerate}

A natural question is therefore: can we train a single compact head over two heavyweight teachers that (a) matches teacher-level accuracy, (b) adds only negligible marginal cost on top of a teacher forward pass that already runs in the analysis chain, and (c) treats the teacher's mistakes as information rather than as ground truth?

This work answers these questions affirmatively, within the bounds set by the teachers themselves. We present \textbf{JetCoRD} (Jet \emph{Co}rrective \emph{R}epresentation \emph{D}istillation), an 82\,k-parameter shared-backbone student trained jointly on two teachers --- GN2-type (ATLAS, 3 classes) and ParT (JetClass, 10 classes) --- and introduce two coupled innovations:

\begin{itemize}
\item \textbf{Adaptive Corrective Representation Distillation (A-CoRD)} is a training-time algorithm that (i) reweights the distillation loss by a per-sample teacher reliability $r_i \in [0,1]$, (ii) repairs erroneous teacher predictions via per-class EMA (exponential moving average) prototypes maintained in the student's embedding space, and (iii) controls the strength of this repair through per-class learnable coefficients $\beta_c$ that contract with teacher class accuracy.
\item \textbf{Reliability-Aware Inference (RAI) fusion} with a \textbf{Class-Conditional Backbone (CCB)} is the architectural counterpart. At inference, a small gate head produces $g_i \in [0,1]$ such that the final logits are a convex mixture $g_i \mathbf{t}_i + (1-g_i)\mathbf{s}_i$ of teacher and student; meanwhile, a soft class hint derived from the teacher's softmax is added to the backbone input. Crucially, $g_i$ is anchored to $r_i$ during training, so the same notion of ``trust the teacher'' applies at train and test time.
\end{itemize}

We summarise our contributions:

\begin{enumerate}
\item \textbf{One shared student for two experiment-specific teachers.} We train a single 82\,k-parameter student jointly over two teachers with different detector responses and disjoint class spaces --- GN2 and ParT. The student \emph{reproduces} both teachers at $\geq\!99\%$ agreement and matches their overall accuracy; the reliability gate then \emph{redistributes} accuracy toward the teachers' weakest classes, raising the JetClass per-class accuracy of $H\!\to\!4q$, QCD and $H\!\to\!bb$ by $+0.80$, $+0.38$ and $+0.24$ percentage points ($|z|>7$) at the cost of comparable drops on $H\!\to\!gg$, $W\!\to\!q\bar q$ and $Z\!\to\!q\bar q$, with overall accuracy unchanged. Every quantity carries binomial, Poisson, and paired-quantity uncertainties; the working points whose rejection deltas fall inside the combined uncertainty are reported as such rather than as gains (Sec.~5). The sharing here is of the \emph{training procedure}, not of a shared representation: the single-domain controls and pre-registered probes of Sec.~6.3 show the two domains remain representationally separate, and we claim no cross-detector transfer.
\item \textbf{A reliability-coupled training/inference framework.} A single per-sample reliability signal $r_i$ controls three mechanisms simultaneously: (i) it down-weights the KL distillation loss on teacher mistakes, (ii) it gates an EMA-prototype repair term that pulls the student toward the true class, and (iii) it anchors the inference-time gate $g_i$ that mixes teacher and student logits. The oracle-assumption KD methods reviewed in Sec.~2 use comparable reliability signals only during training.
\item \textbf{Class-adaptive corrective distillation.} The repair strength is controlled by per-class learnable coefficients $\beta_c$ that automatically contract with the teacher's per-class accuracy. The optimization discovers, without manual tuning, that the repair concentrates on a single class per domain --- $c$ in ATLAS ($\beta_c\approx0.50$) and $H\!\to\!4q$ in JetClass ($\beta_{H4q}\approx0.45$) --- with every other class below $0.20$ (Sec.~7.2).
\item \textbf{A measured ceiling and a decision-theoretic evaluation.} Rather than assert that the correction is bounded, we measure the bound: oracle heads fitted on the frozen teacher embedding --- and on a raw-input side channel and a full per-particle flattening --- recover a budget $B-T$ consistent with zero in both domains, so the correction redistributes errors rather than inflating accuracy (Sec.~7.5). We then recast the gate's value in the terms that matter for deployment --- coverage at a distribution-free certified false-accept rate --- via conformal risk control, and report explicitly where it helps (the poorly-calibrated ATLAS teacher) and where it does not (the already-calibrated ParT teacher) (Sec.~7.4). This, together with the corrected re-evaluation of every table after two evaluation bugs were found and fixed, is the evaluation-rigour contribution of this work.
\end{enumerate}

The remainder is organised as follows. Section~2 reviews related work. Section~3 describes A-CoRD and the JetCoRD architecture. Section~4 details the experimental setup. Sections~5--6 present results and ablations. Section~7 analyses the gate--reliability coupling and the learned $\beta_c$. Section~8 discusses limitations.

\section{Related Work}
\textbf{Jet tagging with deep learning.} Modern jet taggers exploit increasingly detailed substructure information, evolving from BDTs on high-level kinematic features to ParticleNet ~\cite{ParticleNet}, ABCNet ~\cite{ABCNet}, LorentzNet ~\cite{LorentzNet}, PELICAN ~\cite{PELICAN}, and Particle Transformer (ParT) ~\cite{ParT}; state-of-the-art models on the JetClass benchmark achieve $86\,\%$ overall accuracy across ten physics classes with $\mathcal{O}(2\text{M})$ parameters. In ATLAS, GN2 ~\cite{GN2} reaches a $b$-tagging working point of $\varepsilon_b = 0.77$ at a $c$-jet rejection of order $20$, operating on charged-particle graphs together with secondary-vertex information. More recent architectures push further along three axes: (i) \emph{more interaction terms} --- MIParT ~\cite{MIParT} reports a $10\%$ background-rejection improvement on JetClass over ParT by enriching the pair-wise interaction matrix; (ii) \emph{strict symmetry} --- L-GATr ~\cite{LGATr} uses Lorentz-equivariant geometric algebra transformers, and PELICAN ~\cite{PELICAN} achieves explainable permutation and Lorentz-equivariance at $\mathcal{O}(0.1\text{M})$ parameters on the top-tagging benchmark; (iii) \emph{task-agnostic pre-training} --- OmniLearn ~\cite{OmniLearn}, OmniJet-$\alpha$ ~\cite{OmniJet}, MPM ~\cite{MPM}, and Sophon ~\cite{Sophon} train a single foundation backbone that is then fine-tuned for jet tagging, jet generation, and unfolding at $5$--$30\text{M}$ parameters. All of these works push the frontier upward: bigger model, marginally better accuracy. Our work is orthogonal: we ask whether a small correction layer on top of two frozen pre-trained taggers from different experiments can hold their accuracy while serving both from one network.

\textbf{Knowledge distillation in HEP.} Knowledge distillation ~\cite{Hinton} has been applied sporadically in HEP, primarily for inference acceleration: Duarte et al. ~\cite{ABCNetDistill} demonstrated FPGA-friendly compressed jet classifiers; subsequent work in CMS and ATLAS has used KD to reduce HLT-level taggers to deployment-ready sizes. These works treat the teacher as a near-oracle and target a fixed deployment environment. This work, by contrast, explicitly handles teacher mistakes at the sample level and distills \emph{across} experiments. The foundation-model line of work above ~\cite{OmniLearn,OmniJet,MPM,Sophon} in some sense replaces distillation with self-supervised pre-training on large unlabelled jet datasets, but does not address the deployment-side question of compressing the pre-trained backbone into a few-tens-of-thousands-parameter student; the recent vision-domain KD methods discussed below (~\cite{LogitStd,ScaleKD,DKD,SDD,CTKD,WTTM,DIST,DOT,VanillaKD}) have not yet been adapted to HEP taggers.

\textbf{Reliability-aware distillation.} Outside HEP, several works have observed that hard distillation targets are noisy: TAKD ~\cite{TAKD} introduces an intermediate teacher, KDCL ~\cite{KDCL} uses logit averaging over multiple teachers, and RKD ~\cite{RKD} distills relational structure. Menon et al. ~\cite{CAD} provide a statistical framework showing that down-weighting unreliable teacher samples is provably equivalent to a Bayes-optimal correction, while Stanton et al. ~\cite{Stanton} empirically document that vanilla KD systematically fails to recover teacher accuracy when the teacher is imperfect --- a finding directly consistent with our KD baseline (Sec. 5--6). The 2023--2025 KD literature has moved aggressively along the mechanics of the KL loss but still stays inside the ``teacher is an oracle'' frame: DKD ~\cite{DKD} decomposes the KL loss into target-class and non-target-class components and re-weights them statically; SDD ~\cite{SDD} further decouples the loss across spatial scales; Logit Standardization ~\cite{LogitStd} applies a learnable affine rescaling to teacher and student logits to remove the temperature/magnitude mismatch; CTKD ~\cite{CTKD} introduces an \emph{adversarially-scheduled} temperature; WTTM ~\cite{WTTM} proves that the optimal student matches a power-transformed teacher rather than the raw one; DIST ~\cite{DIST} uses Pearson correlation instead of KL to handle stronger teachers; DOT ~\cite{DOT} splits the optimiser between CE and KD branches. Two complementary lines argue from a different direction: VanillaKD ~\cite{VanillaKD} shows on ImageNet that a careful baseline still beats most recent decoupling schemes, while ScaleKD ~\cite{ScaleKD} addresses the teacher-larger-than-student capacity gap with depth/width projectors. \textbf{All of these are oracle-assumption methods} --- they handle the mechanics of the KL loss (or the loss-vs-optimiser interface) but not the content of the teacher's mistakes, and none of them re-uses any of their training-time signals at inference. We extend the reliability-aware line in two ways: (i) our weight $r_i$ explicitly zeros out the KL contribution on teacher mistakes via an indicator on label agreement multiplied by a soft margin gate, and gates an EMA-prototype repair term so the student can override the teacher on classes where the teacher is statistically weak; and (ii) we couple the same $r_i$ to inference via the RAI gate --- a coupling absent from the KD works reviewed above. The closest design in spirit, DOT ~\cite{DOT}, also separates training signals by trust source but does so at the optimiser level only and not at inference.

\textbf{Inference-time teacher mixing.} ``Trust-region'' routing in mixture-of-experts ~\cite{MoE} ~\cite{Switch} makes a discrete choice over experts; the Born-Again Network framework ~\cite{TeacherStudentMix} showed that re-using teacher outputs as soft targets for a same-architecture student yields measurable improvements on image classification, although without a coupled training-time reliability anchor. Our RAI gate departs from both: it is a continuous per-sample mix between teacher and student logits whose value is learned in the same loop as the distillation weight.

\textbf{Multi-domain and cross-detector physics models.} Multi-domain learning in HEP has been explored for simulation-based inference and theory--detector correspondence ~\cite{TheoryDetector}, and for matching observables across collider settings (e.g. OmniFold-style unfolding ~\cite{CrossEnergy}); we instead jointly distill from two different experimental teachers (with disjoint detector responses and disjoint class spaces). The foundation-model line above ~\cite{OmniLearn}~\cite{OmniJet} approaches the cross-task question from the opposite direction (one self-supervised backbone, many downstream heads, one detector simulation); we approach it from the deployment side (two pre-trained detector-specific teachers, one tiny shared student, two detector domains held jointly). What we share is the \emph{training procedure} --- one backbone and one loss trained jointly over both teachers --- not a representation space: no intervention we test (Sec.~6.3) produces a shared representation, and we make no cross-detector genericity claim.

\section{Method}
We propose \textbf{JetCoRD}, a single jet-tagging student with only 82\,k parameters, obtained by distilling two heavyweight teachers --- GN2 and ParT --- into one shared backbone. The student is roughly $20\times$ smaller than either teacher; the design rests on two coupled innovations:

\begin{enumerate}
\item \textbf{Adaptive Corrective Representation Distillation (A-CoRD)} --- a training strategy that adaptively weights teacher knowledge by per-sample reliability and corrects teacher mistakes using learnable class prototypes (Sec. 3.2);
\item \textbf{Reliability-Aware Inference fusion (RAI)} combined with a \textbf{Class-Conditional Backbone (CCB)} --- an architecture-level mechanism that exposes the teacher's soft prediction to both the backbone and the output mixing stage (Sec. 3.3).
\end{enumerate}

The two innovations are tightly coupled: A-CoRD's training-time reliability signal $r_i$ becomes the \emph{anchor} for the inference-time gate in RAI, so the training objective and the deployed network share a consistent notion of ``when to trust the teacher.''

\subsection{Problem Setup and Notation}

We are given two pretrained teachers $\mathcal{T}^{(A)}$ (GN2, ATLAS, $C_A=3$ flavor classes) and $\mathcal{T}^{(J)}$ (ParT, JetClass, $C_J=10$ classes). For every jet $\mathbf{x}_i$ from either dataset we cache:
\begin{itemize}
\item the teacher's jet-level embedding $\mathbf{e}^{T}_i \in \mathbb{R}^{128}$,
\item the teacher logits $\mathbf{t}_i \in \mathbb{R}^{C}$, and
\item the ground-truth label $y_i$.
\end{itemize}

The student maps $(\mathbf{e}^T_i, \text{domain}, \mathbf{t}_i) \to \mathbf{s}_i \in \mathbb{R}^{C}$ through a single shared backbone with two lightweight per-domain heads. The student never sees raw particle-level inputs --- it operates on the teacher's pre-classifier embedding and logits, which is what makes the 82\,k parameter budget feasible. At inference, the teacher must be run once to produce these inputs; the student then refines them. In analysis workflows where teacher outputs are already produced as part of standard data processing, the student adds negligible overhead.

\textbf{Why this is distillation, and how far it can go.} Standard knowledge distillation uses teacher logits as soft targets~\cite{Hinton}; we go further by consuming the teacher's compressed representation $\mathbf{e}^T_i$ as input, using the teacher logits $\mathbf{t}_i$ as both training targets and inference-time fusion partners. A student trained only to match the teacher's soft labels can at best reproduce it; a student trained on the ground-truth labels $y_i$, as ours is, can in principle exceed a teacher whose head is suboptimal on some classes. How far it can go is bounded by the teacher's own errors: the ground-truth cross-entropy term and the EMA-prototype repair act only where the teacher is wrong, so the student tracks the teacher wherever the teacher is correct and deviates only where labeled evidence supports it. The mechanism --- reliability-weighted correction via A-CoRD and gate-anchored inference via RAI --- is the subject of the next sections.

\subsection{Adaptive Corrective Representation Distillation (A-CoRD)}

\subsubsection{Per-sample reliability}

Let $\hat{y}^T_i = \arg\max \mathbf{t}_i$ be the teacher's prediction. We define the per-sample reliability as
\begin{equation}
r_i \;=\; \mathbb{1}\bigl[\hat{y}^T_i = y_i\bigr] \cdot \sigma(\kappa \cdot m_i),
\qquad
m_i = \mathrm{top}_1(\mathbf{t}_i) - \mathrm{top}_2(\mathbf{t}_i),
\end{equation}
where $\sigma(\cdot)$ is the logistic and $\kappa$ controls the sharpness of the soft margin gate ($\kappa = 5$ throughout). $r_i = 0$ if the teacher is wrong and $r_i \in (0,1)$ when the teacher is right; the value scales monotonically with the teacher's logit margin. Concretely:
\begin{itemize}
\item $r_i \to 0$ when the teacher is wrong or under-confident,
\item $r_i \to 1$ when the teacher is right and confident.
\end{itemize}

\subsubsection{Reliability-weighted KD}

The standard temperature-scaled KL distillation loss is reweighted per sample:
\begin{equation}
\mathcal{L}_{\text{KD}}^{w}
= \frac{1}{N}\sum_{i=1}^{N} r_i \cdot T^2 \cdot
\mathrm{KL}\bigl(\sigma_{T}(\mathbf{t}_i)\,\big\|\,\sigma_{T}(\mathbf{s}_i)\bigr),
\end{equation}
so unreliable teacher samples (wrong or low-margin) contribute proportionally less to the gradient. This resolves a well-documented limitation of vanilla knowledge distillation: when the teacher is fallible, forcing the student to mimic its mispredictions actively degrades student performance.

\subsubsection{Class-adaptive repair coefficient}

To repair the cases where the teacher is wrong, A-CoRD additionally uses a representation repair loss based on per-class EMA prototypes $\boldsymbol{\mu}_c$ maintained in the student's penultimate space $\mathbf{z}_i$:
\begin{equation}
\mathcal{L}_{\text{repair}} \;=\; \frac{1}{N}\sum_{i=1}^{N} (1 - r_i)\,\bigl\|\,\mathbf{z}_i - \boldsymbol{\mu}_{y_i}\bigr\|_2^2 .
\end{equation}
This repair term is a class-conditional center loss~\cite{CenterLoss} with EMA class centres; related prototype-based representation learning appears in prototypical networks~\cite{ProtoNet}.
Crucially, the weight on $\mathcal{L}_{\text{repair}}$ is \textbf{per class} and learnable:
\begin{equation}
\beta_c \;=\; \beta_0 \cdot \sigma\bigl(\alpha_c - s \cdot \mathrm{acc}^T_c\bigr),
\end{equation}
where $\mathrm{acc}^T_c$ is the running teacher accuracy on class $c$, $\beta_0 = 0.5$, $s = 5$, and $\alpha_c$ is a learnable scalar (initialized at 5.0; trained with a dedicated learning rate $25\times$ that of the model). Classes where the teacher is already excellent ($\mathrm{acc}^T_c \to 1$) receive small $\beta_c$ --- the student needn't disagree. Classes where the teacher is poor receive large $\beta_c$, encouraging the student to pull its representation toward the true class prototype.

\subsubsection{Full A-CoRD loss}

\begin{equation}
\mathcal{L}_{\text{A-CoRD}} \;=\;
\mathcal{L}_{\text{CE}}(\mathbf{s}_i, y_i)
+ \lambda_{\text{KD}}\,\mathcal{L}_{\text{KD}}^{w}
+ \sum_{c=1}^{C} \beta_c \cdot \mathcal{L}_{\text{repair}}^{(c)}.
\end{equation}

\subsection{Network Architecture: JetCoRD}

The student is a small residual-MLP trunk (two LayerScale ~\cite{LayerScale} ResMLP blocks at $d=64$, FFN multiplier 3) preceded by an input projection and followed by per-domain correction heads (Fig.~\hyperlink{fig:1}{1}). Teacher embedding $\mathbf{e}^T$ and logits $\mathbf{t}$ are taken as input. $\mathbf{e}^T$ is projected to 64\,dim, combined with the CCB class hint (softmax-weighted class embedding), and passed through two ResMLP blocks with LayerScale and a final LayerNorm. The resulting representation $\mathbf{z}$ feeds two heads: the student head produces a logit-space correction $\boldsymbol{\Delta}$, and the gate head produces the per-class reliability estimate $g$. The student logits are written in correction form, $\mathbf{s}^{\mathrm{student}}=\mathbf{t}+\boldsymbol{\Delta}$, so the final output is the convex mixture $g\mathbf{t}+(1-g)\mathbf{s}^{\mathrm{student}}=\mathbf{t}+(1-g)\boldsymbol{\Delta}$. Because the final linear layer of the correction head is zero-initialised ($\boldsymbol{\Delta}\equiv\mathbf{0}$ at step~0), the network is an exact copy of the teacher at initialisation.

\begin{figure}[t]
\centering
\includegraphics[width=0.98\linewidth]{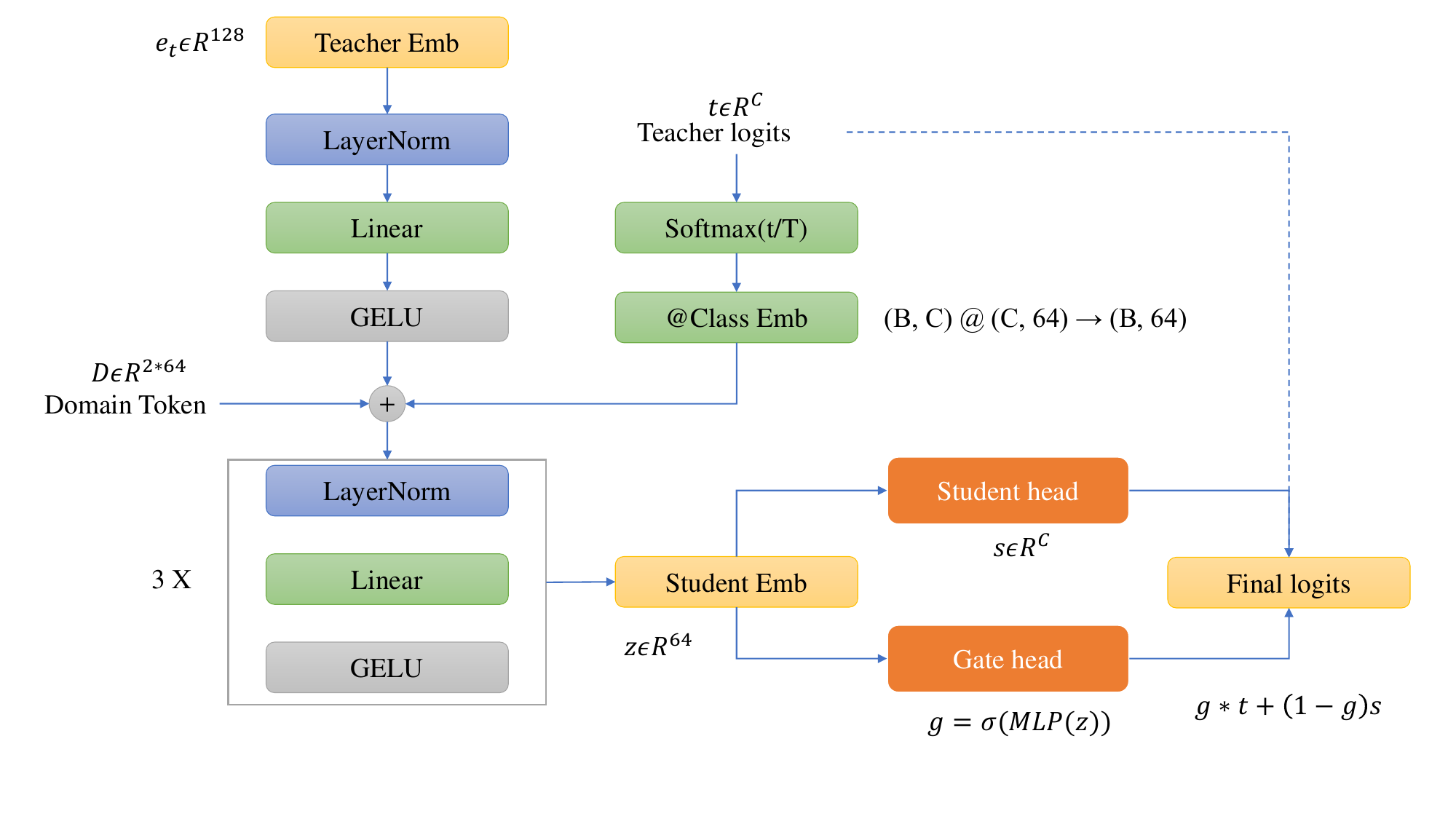}
\hypertarget{fig:1}{}
\caption{JetCoRD architecture.}
\end{figure}

\subsubsection{Class-Conditional Backbone (CCB)}

A naive shared backbone is class-agnostic. We inject a soft class hint derived from the teacher's softmax distribution before the backbone blocks:
\begin{equation}
\mathbf{h}_i \;=\; \mathrm{softmax}(\mathbf{t}_i / T_{\text{CCB}}) \cdot \mathbf{E}^{(d)},
\qquad
\mathbf{z}^{(0)}_i \leftarrow \mathbf{z}^{(0)}_i + \mathbf{h}_i,
\end{equation}
where $\mathbf{E}^{(d)} \in \mathbb{R}^{C_d \times 64}$ is a per-domain class embedding table (zero-initialised, so the network starts identical to a class-agnostic baseline). $T_{\text{CCB}} = 1$ is used as default; we ablate other choices in Sec. 4. Soft mixing avoids the discrete branching pathology of hard mixture-of-experts and allows gradient flow.

\subsubsection{Reliability-Aware Inference Fusion (RAI)}

A per-sample reliability scalar is learned by a small head on the backbone output:
\begin{equation}
g_i \;=\; \sigma\bigl(\mathrm{MLP}_{\text{gate}}(\mathbf{z}_i)\bigr) \in [0,1].
\end{equation}
The final logits are a convex mixture:
\begin{equation}
\mathbf{s}^{\text{final}}_i \;=\; g_i \cdot \mathbf{t}_i + (1 - g_i) \cdot \mathbf{s}^{\text{student}}_i.
\end{equation}
The student logits are written in correction form, $\mathbf{s}^{\text{student}}_i = \mathbf{t}_i + \boldsymbol{\Delta}_i$, where $\boldsymbol{\Delta}_i$ is the per-domain correction produced by the student head; substituting gives $\mathbf{s}^{\text{final}}_i = \mathbf{t}_i + (1 - g_i)\boldsymbol{\Delta}_i$. The final linear layer of the correction head is zero-initialised, so $\boldsymbol{\Delta}_i \equiv \mathbf{0}$ at initialisation and $\mathbf{s}^{\text{final}}_i = \mathbf{t}_i$ \emph{exactly} for every input --- the step-0 invariant invoked throughout this paper --- while the gate bias is initialised to $+2.5$ so that training starts from a high-trust regime ($g_i \approx 0.92$) and the correction is grown only as the data support it.
$g_i \to 1$ means ``trust teacher'' ($\mathbf{s}^{\text{final}}_i \to \mathbf{t}_i$), $g_i \to 0$ means ``apply the full correction''. To prevent the trivial $g_i \equiv 1$ solution that would collapse the correction, we introduce two auxiliary losses:
\begin{itemize}
\item a \textbf{student CE auxiliary} $\mathcal{L}_{\text{stu-CE}} = \mathrm{CE}(\mathbf{s}^{\text{student}}_i, y_i)$ with weight $w_{\text{stu}}=0.5$, keeping the student head's own logits label-discriminative so the un-gated model (Sec.~6.4) stays usable, and
\item a \textbf{gate anchor}:
\end{itemize}
\begin{equation}
\mathcal{L}_{\text{gate}} \;=\; \mathrm{BCE}\bigl(\mathrm{logit}(g_i),\, r_i\bigr),\qquad w_{\text{gate}}=0.03,
\end{equation}
which ties the inference-time gate to the A-CoRD training-time reliability $r_i$. Because $r_i$ is by construction an agreement indicator times a margin factor, this anchor makes $g_i$ a \emph{learned proxy for teacher correctness}: it is trained to estimate, from the backbone representation, the probability that the teacher is right on this jet.

\subsubsection{Full training objective}

For each domain $d \in \{A, J\}$:
\begin{equation}
\mathcal{L}^{(d)} \;=\;
\underbrace{\mathrm{CE}(\mathbf{s}^{\text{final}}_i, y_i)}_{\text{main task}}
+ \mathcal{L}_{\text{A-CoRD}}^{(d)}(\mathbf{s}^{\text{student}}_i)
+ w_{\text{stu}}\,\mathcal{L}_{\text{stu-CE}}^{(d)}
+ w_{\text{gate}}\,\mathcal{L}_{\text{gate}}^{(d)} .
\end{equation}
The two-domain total is $\mathcal{L} = \mathcal{L}^{(A)} + \mathcal{L}^{(J)}$, optimised in a single micro-step.

\subsection{Algorithm-Architecture Coupling}

The defining feature of JetCoRD is that A-CoRD (training algorithm) and RAI/CCB (inference architecture) share the same reliability concept (\hyperlink{tab:1}{Table~1}):

\hypertarget{tab:1}{}
\textbf{Table 1. Algorithm--architecture coupling.} The A-CoRD training algorithm and the RAI/CCB inference architecture are two views of one per-sample reliability $r_i$; this shared concept is what makes the design interpretable end to end.

\begin{center}
\begin{tabular}{l p{9.2cm}}
\toprule
Where it appears & What it does \\
\midrule
A-CoRD KD weight $r_i$ & Down-weights distillation gradient on unreliable teacher samples \\
A-CoRD repair $1-r_i$ & Pulls student representation toward true class prototype \\
RAI gate $g_i$ & Mixes teacher and student logits at inference \\
Gate anchor BCE($g_i$, $r_i$) & Ties inference gate to the A-CoRD reliability \\
CCB class hint & Injects teacher's confidence pattern into the backbone \\
\bottomrule
\end{tabular}
\end{center}
This coupling provides interpretability for free: a sample with high $r_i$ during training will also have high $g_i$ at inference and the student is essentially copying the teacher; a sample with low $r_i$ trains the student's own path and tells the gate to use it.

\section{Experimental Setup}
\subsection{Datasets and Teachers}

\textbf{ATLAS (3 classes).} The official GN2 weights and the full ATLAS training sample are not public. We therefore train our own ATLAS teacher --- a GN2-type track-based flavour tagger ($\sim\!1.55\text{M}$ parameters) built with the public Salt framework ~\cite{Salt} --- on the publicly released $\mathrm{t}\bar{\mathrm{t}}$ Delphes fast-simulation sample of Ref.~\cite{atlasdata}, using the same charged-particle inputs and $b$/$c$/light labelling as GN2 ~\cite{GN2}. Because the teacher is trained on fast simulation (and not on calibrated collision data), its absolute background-rejection performance lies a factor of $\mathcal{O}(2)$ below the published GN2 curves; all \emph{relative} student-versus-teacher comparisons in this work are therefore internally consistent, but the absolute student rejection numbers should not be read as GN2-grade performance. Classes are $b$, $c$, light ($u$), with proportions approximately $1\!:\!1\!:\!4$ before reweighting. We cache the teacher's 128-dimensional jet embedding (taken from the layer immediately before the classification head) and 3-class logits for every jet in train, validation, and test splits. The test split contains $1.35\text{M}$ jets.

\textbf{JetClass (10 classes).} We use the official JetClass test split processed through the public ParT ~\cite{ParT} checkpoint ($\sim 2\text{M}$ parameters). Classes are QCD, $H\!\to\!bb$, $H\!\to\!cc$, $H\!\to\!gg$, $H\!\to\!4q$, $H\!\to\!q\bar q\ell$, $Z\!\to\!q\bar q$, $W\!\to\!q\bar q$, $t\!\to\!bq\bar q$, $t\!\to\!b\ell\nu$. The test split contains $20\text{M}$ jets.

Both teachers share an output embedding dimensionality of 128, which we use as the student's input. Both caches are stored in HDF5 with \texttt{(N, 128)} float32 teacher embeddings and \texttt{(N, C)} float32 teacher logits.

\subsection{Student Configuration}

The default JetCoRD student uses:
\begin{itemize}
\item input projection $128 \to 64$,
\item $L = 2$ ResMLP blocks at $d=64$ with FFN multiplier 3 and LayerScale init $\gamma_0 = 10^{-4}$,
\item two per-domain correction heads ($64 \to 128 \to C_d$; final linear layer zero-initialised, so $\boldsymbol{\Delta}\equiv\mathbf{0}$ at step~0),
\item one shared RAI gate head ($64 \to 64 \to 1$),
\item per-domain CCB class embeddings $C_d \times 64$ initialised to zero,
\item domain token embedding (2 $\times$ 64) initialised to zero.
\end{itemize}

Total parameter count: \textbf{82,234}.

\subsection{Training Protocol}

We train for 40 epochs at batch size 8192 on a single NVIDIA L40 GPU with mixed precision (\texttt{amp=true}). Optimiser settings:
\begin{itemize}
\item AdamW, learning rate $2 \times 10^{-3}$, weight decay $10^{-5}$, warm-up 800 steps;
\item a separate SGD optimiser for the A-CoRD coefficients $\{\alpha_c\}$ at learning rate 0.05, initialised at $\alpha_c = 5.0$;
\item exponential moving average of model weights with decay 0.999 (used at inference);
\item early stopping on a combined validation metric (overall ACC + 0.05 $\times$ normalised bg-rejection at key working points; see Sec. 4.5) with patience 25 epochs.
\end{itemize}

\subsection{Default A-CoRD and RAI Hyper-parameters}

Determined by a ten-round random search (Sec. 6.5; \hyperlink{tab:2}{Table~2}):

\hypertarget{tab:2}{}
\textbf{Table 2. Default A-CoRD and RAI hyper-parameters.} Values selected by a ten-round random search (Sec.~6.5).

\begin{center}
\begin{tabular}{lcc}
\toprule
Hyper-parameter & Value & Description \\
\midrule
$\beta_0$ & 0.5 & base repair-loss scale \\
$s$ (acc scale) & 5.0 & shrinks $\beta_c$ as teacher class-accuracy grows \\
$\kappa$ & 5.0 & margin sharpness in reliability $r_i$ \\
$T_\mathrm{KD}$ & 4.0 & KD temperature \\
$\lambda_\mathrm{KD}$ & 1.0 & KD loss weight \\
EMA momentum (proto) & 0.99 & per-class prototype EMA \\
proto warm-up steps & 200 & freeze prototypes during first 200 steps \\
$w_\mathrm{stu}$ & 0.5 & auxiliary student-CE weight \\
$w_\mathrm{gate}$ & 0.03 & gate-anchor BCE weight\\
$T_\mathrm{CCB}$ & 1.0 & CCB softmax temperature \\
gate bias init & $+2.5$ & initial gate $g_i \approx 0.92$ \\
\bottomrule
\end{tabular}
\end{center}
\subsection{Evaluation Metrics}

Following the conventions in ~\cite{GN2} and ~\cite{ParT} we report two families of metrics on the held-out test split:

\begin{enumerate}
\item \textbf{Overall and per-class accuracy.} Reported as percentages.
\item \textbf{Background rejection $1/\varepsilon_b$} at a target signal efficiency. For ATLAS we report $b$-vs-$u$, $b$-vs-$c$ at two working points each and $c$-vs-$u$, $c$-vs-$b$ at three. For JetClass we report the seven medium-purity rejections at $\varepsilon = 0.5$ together with the two high-purity points used by ParT: $H\!\to\!q\bar q\ell$ at $\varepsilon = 0.99$ and $t\!\to\!b\ell\nu$ at $\varepsilon = 0.995$.
\end{enumerate}

\textbf{Precise definition of $1/\varepsilon_b$.} Throughout this paper we adopt the \textbf{one-vs-one (OvO) subset} definition that is standard in the ATLAS flavour-tagging publications (e.g. GN2 ~\cite{GN2}) and in the weaver/ParT ~\cite{ParT} codebase. For a chosen signal class $s$ and background class $b$, we restrict the test set to jets with true label $y \in \{s, b\}$ and compute the per-jet binary discriminant
\begin{equation}
D_i \;=\; \frac{P_s(\mathbf{x}_i)}{P_s(\mathbf{x}_i) + P_b(\mathbf{x}_i)} \;\in\; [0,1],
\end{equation}
which is the (softmax-renormalised) binary likelihood ratio between the two classes. We set the threshold to the $(1{-}\varepsilon_s)$-quantile of $\{D_i : y_i = s\}$ and report
\begin{equation}
\varepsilon_b \;=\; \frac{|\{i : y_i = b,\ D_i \ge \tau\}|}{|\{i : y_i = b\}|},
\qquad
1/\varepsilon_b \;\text{(``background rejection'').}
\end{equation}
For ATLAS flavour-tagging we additionally use the GN2 log-likelihood discriminant definitions $D_b = \log P_b/(f_c P_c + (1-f_c) P_u)$ with $f_c = 0.20$ and the analogous $D_c$ with $f_b = 0.30$, which are monotonically equivalent to the binary form for the two-class regime and reproduce the published ATLAS working-point definitions. Our teacher row is obtained by evaluating the official ParT checkpoint (\texttt{ParT\_full.pt}) on the full 20M JetClass test set.

For model selection during training, we use a combined metric: overall ACC $+$ 0.05 $\times$ normalised bg-rejection at 12 representative working points (6 per domain). This avoids saturation of the plain-ACC metric, which plateaus after the first epoch.

\subsection{Multi-seed Protocol}
All summary statistics in Sections 5--6 are reported as mean $\pm$ standard deviation over $n=10$ random seeds. Seeds change the PyTorch RNG (init + dropout) but not the data ordering, since data caches are deterministic.

\section{Results}
We evaluate the JetCoRD framework with four algorithm configurations sharing the same backbone (RAI + CCB): \textbf{KD} --- vanilla KL distillation, no reliability weighting, no prototype repair; \textbf{CoRD} --- reliability-weighted KD + prototype repair with fixed $\beta_c = 0.5$. \textbf{A-CoRD} --- reliability-weighted KD + prototype repair with learnable per-class $\beta_c$ + gate-anchor coupling. \textbf{PCG} --- A-CoRD with per-class RAI gate (default JetCoRD configuration). JC results use the full 20M JetClass test set; ATLAS uses 1.35M. $\Delta\% = (s-t)/t \times 100$. The overall-accuracy comparison (\hyperlink{tab:3}{Table~3}) is a tie within the seed-to-seed uncertainty and is left unbolded; in the remaining tables \textbf{bold} marks the better entry of each comparison.

\subsection{Overall Accuracy}

\hypertarget{tab:3}{}
\textbf{Table 3. Overall accuracy (\%), all four configurations.} Mean $\pm$ std over $n=10$ seeds; the four configurations share one backbone (RAI + CCB). Every difference from the teacher is within the seed-to-seed uncertainty, so no entry is bolded.

\begin{center}
\begin{tabular}{lccc}
\toprule
Method  & ATLAS ACC & JC ACC \\
\midrule
Teacher  & 73.932 \% & 86.051 \% \\
KD  & 73.932 $\pm$ 0.001 \% & 86.051 $\pm$ 0.000 \% \\
CoRD  & 73.927 $\pm$ 0.013 \% & 86.057 $\pm$ 0.005 \% \\
A-CoRD  & 73.938 $\pm$ 0.012 \% & 86.062 $\pm$ 0.004 \% \\
PCG  & 73.935 $\pm$ 0.019 \% & 86.064 $\pm$ 0.005 \% \\
\bottomrule
\end{tabular}
\end{center}
Every configuration reproduces the teacher's overall accuracy to better than $0.02\%$; the student is not a compression win on this axis, and we do not claim one. KD is a bit-exact copy by construction (its std and paired SE are at the $10^{-3}\%$ level). The reliability-aware configurations (CoRD, A-CoRD, PCG) move by at most $1\sigma$ of the seed-to-seed std, which is why the rest of the paper reports per-working-point and per-class behaviour rather than overall accuracy.
\subsection{Background Rejection}

\hypertarget{tab:4}{}
\textbf{Table 4. JetCoRD (PCG) vs Teacher on all 19 working points.} ATLAS uses the GN2 $D_b$/$D_c$ discriminants; JC uses OvO $D = P_s/(P_s+P_b)$. The PCG column is the mean $\pm$ seed-to-seed std over the $n=10$ seeds. $\Delta\%$ = PCG $-$ Teacher; the uncertainty quoted on $\Delta\%$ is the \emph{combined} per-seed uncertainty, $\sigma_\Delta=\sqrt{\sigma_\text{count}^2+(s/\sqrt{n})^2}$ --- the paired jet-counting error $\sigma_\text{count}$ on the shared test set folded in quadrature with the seed-to-seed standard error of the mean --- and is dominated by the seed term at most points. Positive $\Delta\%$ = student exceeds teacher; points with $|\Delta|/\sigma_\Delta\ge 2$ are marked $\ast$. The $t\!\to\!bqq$@$0.5$ point is count-limited ($n_\text{pass}<100$) and marked $\dagger$; its percentage should be read as quantization-dominated.

\begin{center}
\begin{tabular}{lccc}
\toprule
Working point & Teacher & PCG (JetCoRD) & $\Delta$\% ($\sigma_\Delta$) \\
\midrule
\multicolumn{4}{c}{\textbf{ATLAS}} \\
$b$-vs-$u$ @0.7  & 492.3  & 494.2 $\pm$ 10.1 & $+0.4 \pm 1.0$ \\
$b$-vs-$u$ @0.77 & 163.3  & 157.9 $\pm$ 4.7  & $-3.3 \pm 1.0^{\ast}$ \\
$b$-vs-$c$ @0.7  & 17.52  & 17.51 $\pm$ 0.04 & $-0.1 \pm 0.1$ \\
$b$-vs-$c$ @0.77 & 8.47   & 8.49 $\pm$ 0.03 & $+0.2 \pm 0.2$ \\
$c$-vs-$u$ @0.2  & 785.3  & 773.7 $\pm$ 23.5 & $-1.5 \pm 1.3$ \\
$c$-vs-$u$ @0.3  & 208.9  & 208.5 $\pm$ 3.5  & $-0.2 \pm 0.7$ \\
$c$-vs-$u$ @0.4  & 67.7   & 67.4 $\pm$ 1.1   & $-0.4 \pm 0.6$ \\
$c$-vs-$b$ @0.2  & 41.5   & 41.6 $\pm$ 0.2 & $+0.2 \pm 0.2$ \\
$c$-vs-$b$ @0.3  & 18.20  & 18.26 $\pm$ 0.08 & $+0.3 \pm 0.2$ \\
$c$-vs-$b$ @0.4  & 9.99   & 10.01 $\pm$ 0.06 & $+0.2 \pm 0.2$ \\
\midrule
\multicolumn{4}{c}{\textbf{JetClass}} \\
Hbb @0.5  & 10638.3 & 10736.2 $\pm$ 101.6 & $+0.9 \pm 1.7$ \\
Hcc @0.5  & 4149.4  & 4191.3 $\pm$ 26.8  & $+1.0 \pm 1.0$ \\
Hgg @0.5  & 123.44  & 123.31 $\pm$ 0.12  & $-0.1 \pm 0.1$ \\
H4q @0.5  & 1869.2  & 1877.6 $\pm$ 20.7  & $+0.5 \pm 0.8$ \\
Tbqq @0.5$^\dagger$ & 32258.1 & 33120.2 $\pm$ 530.7 & $+2.7 \pm 2.5$ \\
Wqq @0.5  & 542.7   & 541.4 $\pm$ 2.6   & $-0.2 \pm 0.4$ \\
Zqq @0.5  & 402.3   & 402.8 $\pm$ 1.2 & $+0.1 \pm 0.4$ \\
Hqql @0.99 & 5420.1 & 5333.1 $\pm$ 85.5 & $-1.6 \pm 1.3$ \\
Tbl @0.995 & 16260.2 & 15789.8 $\pm$ 280.7 & $-2.9 \pm 2.3$ \\
\bottomrule
\end{tabular}
\end{center}
Read by sign alone, PCG exceeds the teacher on 10 of the 19 working points and falls below on 9; but once the combined uncertainty $\sigma_\Delta$ of Table~4 is attached, \emph{every positive $\Delta$ among these 19 is within $1\sigma$ of zero} --- not one is a significant improvement --- while the single entry that reaches $|\Delta|/\sigma_\Delta\ge2$ is a regression, the ATLAS $b$-vs-$u$@0.77 point ($-3.3\%$, $z=3.3$). The sign pattern is nonetheless structured rather than random: the positive entries cluster on the physics-actionable hadronic boundaries (JC $H\!\to\!bb$, $H\!\to\!cc$, $H\!\to\!4q$; ATLAS $c$-vs-$b$), and the seed-to-seed term dominates $\sigma_\Delta$ at almost every point. (For completeness, the wider evaluation sweep also carries two ATLAS points at $\varepsilon_b=0.85$, outside the primary set, that move at $z\simeq2.5$: $b$-vs-$u$ by $+0.9\%$ and $b$-vs-$c$ by $-0.2\%$ --- sign-alternating, quantization-level movements at the extreme tail.) We therefore claim no per-working-point rejection advantage. The statistically robust effect is instead the per-class \emph{redistribution} of Sec.~5.3 (\hyperlink{tab:5}{Table~5}), where the per-class accuracy changes reach $|z|>7$; the multi-method comparison (KD, CoRD, A-CoRD vs PCG) is in \hyperlink{tab:7}{Table~7}, and the direct comparison against simple calibration baselines is in Sec.~7.3.

\textbf{Why do Hqql@0.99 and Tbl@0.995 regress?}
These two working points probe the extreme tail of the discriminant distribution at $\varepsilon_s \ge 0.99$, where the teacher already achieves 98--99\% per-class accuracy. At such high purity, the discriminant threshold $D_\text{threshold} \approx 0.999$ sits on a near-perfect teacher softmax. A-CoRD's prototype repair mechanism perturbs the student's per-class softmax by $\mathcal O(10^{-3})$ in $P_s/(P_s+P_b)$---negligible for overall accuracy but sufficient to displace a handful of background jets across the quantile threshold, multiplying $\varepsilon_b$ by a factor of $\sim$1.02--1.03. This is a structural consequence of repairing teacher mistakes in representation space: the repair that helps the weakly-classified channels unavoidably introduces small logit-space perturbations that are visible only at the most extreme working points. Section~8 discusses this trade-off and two mitigations.

\subsection{Per-Class Accuracy}

\hypertarget{tab:5}{}
\textbf{Table 5. Per-class accuracy (\%), PCG vs Teacher.} Mean $\pm$ std over $n=10$ seeds. $\Delta$ = PCG $-$ Teacher in percentage points. The paired per-jet statistical error on each $\Delta$ is $\le 0.016$\,pp in both domains; it is not the limiting uncertainty --- the seed-to-seed spread (the $\pm$ shown) is larger by an order of magnitude, and it is the seed-to-seed term that sets the per-class $z$ quoted in Sec.~5.3.

\begin{center}
\begin{tabular}{lccc}
\toprule
Class & Teacher & PCG (JetCoRD) & $\Delta$ (pp) \\
\midrule
\multicolumn{4}{c}{\textbf{ATLAS}} \\
$b$ & 76.69 & 76.67 $\pm$ 0.16 & $-0.02$ \\
$c$ & 57.84 & \textbf{58.07} $\pm$ 0.60 & \textbf{+0.23} \\
$u$ & 87.27 & 87.07 $\pm$ 0.50 & $-0.19$ \\
\midrule
\multicolumn{4}{c}{\textbf{JetClass}} \\
QCD & 77.73 & \textbf{78.11} $\pm$ 0.16 & \textbf{+0.38} \\
$H\!\to\!bb$ & 92.52 & \textbf{92.76} $\pm$ 0.08 & \textbf{+0.24} \\
$H\!\to\!cc$ & 84.43 & \textbf{84.45} $\pm$ 0.10 & \textbf{+0.01} \\
$H\!\to\!gg$ & 79.85 & 79.09 $\pm$ 0.23 & $-0.75$ \\
$H\!\to\!4q$ & 84.68 & \textbf{85.48} $\pm$ 0.24 & \textbf{+0.80} \\
$H\!\to\!q\bar q\ell$ & 98.07 & 98.04 $\pm$ 0.02 & $-0.03$ \\
$Z\!\to\!q\bar q$ & 69.39 & 69.12 $\pm$ 0.28 & $-0.27$ \\
$W\!\to\!q\bar q$ & 79.94 & 79.62 $\pm$ 0.24 & $-0.31$ \\
$t\!\to\!bq\bar q$ & 95.35 & \textbf{95.41} $\pm$ 0.05 & \textbf{+0.06} \\
$t\!\to\!b\ell\nu$ & 98.55 & \textbf{98.56} $\pm$ 0.01 & \textbf{+0.01} \\
\bottomrule
\end{tabular}
\end{center}
Across all 13 classes, PCG matches or exceeds the teacher on 7. Unlike the working-point deltas of Table~4, the per-class accuracy changes are statistically resolved: the largest movers run at $|z|>7$ under the seed-to-seed uncertainty --- $H\!\to\!4q$ ($+0.80$\,pp, $z=10.6$), $H\!\to\!bb$ ($+0.24$\,pp, $z=9.4$) and QCD ($+0.38$\,pp, $z=7.6$) up, and $H\!\to\!gg$ ($-0.75$\,pp, $z=-10.3$), $W\!\to\!q\bar q$ ($-0.31$\,pp, $z=-4.1$) and $Z\!\to\!q\bar q$ ($-0.27$\,pp, $z=-3.1$) down. The mean over the 13 classes is zero to within resolution and the overall accuracy is unchanged, so this is a genuine \emph{redistribution} of the teacher's per-class errors rather than an inflation --- precisely what the measured ceiling of Sec.~7.5 predicts. The three ATLAS classes ($b$, $c$, $u$) all stay within $1.3\sigma$ of the teacher, including the ATLAS $c$-jet class (nominal $+0.23$\,pp, teacher accuracy 57.8\,\%). This mixed per-class pattern is consistent with the repair-budget allocation reported in Table~11, and we present it as such rather than as a uniform improvement.

\section{Ablation Studies}
We perform two complementary ablations --- architecture (§6.1) and algorithm (§6.2) --- two controls (single-domain sharing, §6.3; the effect of the inference gate, §6.4), and a hyper-parameter sensitivity scan (§6.5). The algorithm comparison uses $n=10$ seeds; the architecture ablation uses $n=3$, since each variant is a distinct configuration (and must not be pooled), as noted where relevant.

\subsection{Architecture Ablation}

Holding the A-CoRD loss fixed, we remove RAI and CCB individually ($n=3$ for the three reduced variants; the full configuration is the $n=10$ PCG arm; \hyperlink{tab:6}{Table~6}):

\hypertarget{tab:6}{}
\textbf{Table 6. Architecture ablation (A-CoRD loss).} ATLAS and JC overall accuracy and the ATLAS $b$-vs-$c$@0.77 shift; mean $\pm$ seed std. The three reduced variants are $n=3$; the full configuration is $n=10$.

\begin{center}
\begin{tabular}{lccc}
\toprule
Variant  & ATLAS ACC \% & JC ACC \% & $b$-vs-$c$ @0.77 $\Delta$\% \\
\midrule
w/o RAI, w/o CCB  & 73.950 $\pm$ 0.089 & 86.041 $\pm$ 0.017 & $+0.04$ \\
+CCB only  & 73.941 $\pm$ 0.008 & 86.055 $\pm$ 0.006 & $+0.10$ \\
+RAI only  & 73.913 $\pm$ 0.018 & 86.056 $\pm$ 0.001 & $-0.02$ \\
Full (RAI+CCB)  & 73.935 $\pm$ 0.019 & \textbf{86.064} $\pm$ 0.005 & \textbf{+0.22} \\
\bottomrule
\end{tabular}
\end{center}
The straightforward reading is that the architectural components have \emph{small} effects on the primary metrics: all four variants sit within $\sim 1\sigma$ of one another and of the teacher, and no single component produces the large $b$-vs-$c$@0.77 swing that an earlier, mis-evaluated analysis suggested. What the coupling does provide is a monotone ordering in JC accuracy (86.041 without either component $\to$ 86.055/86.056 with one $\to$ 86.064 with both) and the largest positive ATLAS $b$-vs-$c$@0.77 shift ($+0.22\%$, against $+0.04\%$ for neither component and $+0.10\%$/$-0.02\%$ for CCB-only/RAI-only), where the two components act in the same direction. We therefore treat the architecture as a modest, consistent refinement rather than the dominant mechanism; the step-0 structural guarantee (Sec.~3.3) does not depend on it.

\subsection{Algorithm Ablation}

Holding the backbone (RAI + CCB) fixed, we compare loss configurations:

\hypertarget{tab:7}{}
\textbf{Table 7. Algorithm ablation ($n=10$).} JC overall accuracy, representative working-point shifts, and the mean $\Delta\%$ over the full 19-WP set with the number of WPs improved by more than $0.05\%$ (counted by sign; with the combined per-seed uncertainty of Table~4, none of these single working points clears $2\sigma$).

\begin{center}
\small
\begin{tabular}{lccccc}
\toprule
Method & JC ACC \% & $b$-vs-$c$ @0.77 $\Delta$\% & Hbb @0.5 $\Delta$\% & mean $\Delta$\% & WPs $>\!+0.05\%$ \\
\midrule
KD  & 86.051 & $+0.01$ & $+0.05$ & $-0.06$ & 3/19 \\
CoRD  & 86.057 & $-0.16$ & $+0.75$ & $-0.33$ & 8/19 \\
A-CoRD  & 86.062 & $+0.18$ & $+0.32$ & $-0.45$ & 6/19 \\
\textbf{PCG}  & \textbf{86.064} & \textbf{+0.22} & \textbf{+0.92} & $-0.19$ & \textbf{10/19} \\
\bottomrule
\end{tabular}
\end{center}
KD reproduces the teacher closely (mean $\Delta = -0.06\%$ over 19 WPs; 15 of them within $\pm0.05\%$). The reliability-aware variants do not uniformly beat it: they {\em redistribute} the behaviour, nominally gaining on the physics-actionable hadronic boundaries (PCG: $H\!\to\!bb$@0.5 $+0.92 \pm 1.7\%$, ATLAS $b$-vs-$c$@0.77 $+0.22 \pm 0.2\%$, $H\!\to\!cc$ $+1.0 \pm 1.0\%$) while regressing on the saturated semi-leptonic tails and on ATLAS $b$-vs-$u$@0.77 --- though every one of these per-point shifts lies within the combined uncertainty of Table~4, so what survives the uncertainty is the sign pattern and the per-class redistribution of Sec.~5.3, not any single working point. PCG improves the most working points (10/19) and has the best JC accuracy and the largest $H\!\to\!bb$ shift, but its mean $\Delta$ is still negative. We report it as a targeted, not a net, improvement.

\subsection{Single-Domain Control: Does Sharing Cost Anything?}

A natural concern is that forcing one backbone to serve two different experiments degrades both, relative to training a dedicated student per domain. We test this directly: we train two single-domain controls with the same recipe as the joint model, one on ATLAS only and one on JetClass only, each with $n=10$ seeds. \hyperlink{tab:8}{Table~8} reports the outcome.

\hypertarget{tab:8}{}
\textbf{Table 8. Single-domain control ($n=10$).} Overall accuracy of the joint A-CoRD model (recipe- and seed-matched to the single-domain controls) against two dedicated single-domain students and against the frozen teacher. The single-domain controls never update the untouched domain's correction branch, so that column falls back to the teacher; the informative comparison is each domain's own benchmark.

\begin{center}
\begin{tabular}{lcc}
\toprule
Training data  & ATLAS ACC \% & JC ACC \% \\
\midrule
Teacher (frozen)  & 73.93 & 86.05 \\
Joint, A-CoRD (both domains)  & 73.94 $\pm$ 0.01 & \textbf{86.06} $\pm$ 0.00 \\
ATLAS-only  & \textbf{73.95} $\pm$ 0.02 & 86.03 $\pm$ 0.02 \\
JC-only  & 73.84 $\pm$ 0.24 & 86.06 $\pm$ 0.00 \\
\bottomrule
\end{tabular}
\end{center}

On each domain's own benchmark the joint model matches its dedicated single-domain control to within $0.02$\,pp --- far inside the seed-to-seed std. The other-domain column is not a transfer test: the single-domain controls never update the untouched domain's correction branch, so its output falls back to the teacher there --- JC-only leaves ATLAS at $73.84 \pm 0.24$ against the teacher's $73.93$, and ATLAS-only leaves JetClass at $86.03 \pm 0.02$ against the teacher's $86.05$. What the control isolates is therefore the own-domain comparison, which shows that cross-experiment sharing costs nothing measurable: one training run and one 82\,k-parameter backbone serve both teachers as well as two dedicated models would, which is the architectural claim we set out to test.

\subsection{Effect of the Inference Gate}

The inference gate $g_i$ mixes teacher and student logits; setting it to zero ($g_i\equiv 0$) isolates the student head on its own. We re-evaluate every trained checkpoint under this setting. \hyperlink{tab:9}{Table~9} reports overall accuracy together with two representative JetClass rejection working points.

\hypertarget{tab:9}{}
\textbf{Table 9. Gated vs un-gated inference ($n=10$).} Rejection values are $1/\varepsilon$ at the stated working point; the teacher row is the shared reference.

\begin{center}
\begin{tabular}{lcccc}
\toprule
Configuration  & ATLAS ACC \% & JC ACC \% & JC $H\!\to\!bb$@0.5 & JC $H\!\to\!cc$@0.5 \\
\midrule
Teacher  & 73.93 & 86.05 & 10638 & 4149 \\
KD, gate off  & 73.93 $\pm$ 0.01 & 86.03 $\pm$ 0.01 & 9988 & 3595 \\
KD, gate on  & 73.93 $\pm$ 0.00 & 86.05 $\pm$ 0.00 & 10644 & 4150 \\
PCG, gate off  & 73.92 $\pm$ 0.02 & 86.02 $\pm$ 0.03 & 9797 & 3558 \\
PCG, gate on  & 73.94 $\pm$ 0.02 & \textbf{86.06} $\pm$ 0.00 & \textbf{10736} & \textbf{4191} \\
\bottomrule
\end{tabular}
\end{center}

The un-gated student head is a respectable classifier in accuracy terms --- it reproduces the teacher's overall accuracy to within $0.03$\,pp --- but it is a substantially worse \emph{discriminant} at the high-purity tails: with the gate off, the JC $H\!\to\!bb$ and $H\!\to\!cc$ rejections at $\varepsilon=0.5$ fall by $6$--$15\%$ relative to the teacher, whereas with the gate on they recover teacher-level rejection. The reason is that background rejection is set by the extreme right tail of the student softmax, which a small distilled head does not reproduce even when its argmax accuracy does. This is the sharpest argument for the coupled design: the gate is not a marginal add-on to an already-good student --- it is what makes the small head usable at analysis-grade working points. It also answers directly the question of what the independent student head does alone: it does \emph{not} recover teacher-level discrimination, and so the design deliberately keeps the teacher's behaviour as the default, deviating from it only where the reliability signal supports it.

\subsection{Hyper-parameter Sensitivity}
A ten-round random search over 16 configurations identified the gate-anchor weight $w_\text{gate}$ as the most sensitive parameter. At $w_\text{gate} = 0.05$ the gate saturates (collapsing to teacher copy); at $w_\text{gate} = 0.0$ the gate decouples from $r_i$. The optimal value $w_\text{gate} = 0.03$ is used throughout. The KD temperature $T = 4.0$ and prototype EMA momentum 0.99 are robust within $\pm 50\%$ of their nominal values.

\section{Analysis}
The ablations in Sec.~6 show that neither the architecture nor the algorithm produces a large gain in the primary metrics; the value of the design is the reliability coupling itself. This section examines what that coupling actually does. We check three things: (i) the inference gate $g_i$ concentrates on samples where the teacher is reliable, more strongly on ATLAS than on JetClass (Sec.~7.1); (ii) the learnable per-class $\beta_c$ converges to a sparse repair pattern tied to the teacher's weak spots (Sec.~7.2); and (iii) whether the gate beats simple, label-free reliability baselines --- it does not, uniformly, and we say so (Sec.~7.3).

\subsection{Gate--Reliability Coupling}

We evaluate the trained PCG checkpoint on the full test split and record both the inference-time gate $g_i$ and the training-style reliability $r_i$ (computed analytically from teacher logits and the ground-truth label, but only used as a probe here, not as a training signal). \hyperlink{tab:10}{Table~10} summarises the conditional statistics; \hyperlink{fig:2}{Fig.~2} visualises the distribution.

\hypertarget{tab:10}{}
\textbf{Table 10. Inference gate vs teacher correctness.} Quantities measured on the full test split for the PCG configuration (representative seed). $\bar r$ and $\bar g$ are the test-set means; $\mathrm{AUC}(g)$ is the AUC of $g$ for ``teacher predicted this jet correctly'' (higher = better); it carries a Hanley--McNeil standard error of $4\times10^{-4}$ (ATLAS) and $1\times10^{-4}$ (JetClass).

\begin{center}
\begin{tabular}{lcc}
\toprule
Quantity & ATLAS & JC \\
\midrule
Test-set size $N$ & 1.35 M & 20.00 M \\
Teacher accuracy & 73.93 \% & 86.05 \% \\
Mean reliability $\bar r$ & 0.712 & 0.850 \\
Mean inference gate $\bar g$ & 0.717 & 0.981 \\
Pearson $\mathrm{corr}(g,r)$ & 0.50 & 0.39 \\
$\langle g \mid \hat y^T = y \rangle$ (correct) & 0.768 & 0.986 \\
$\langle g \mid \hat y^T \ne y \rangle$ (wrong) & 0.572 & 0.952 \\
$\mathrm{AUC}(g)$ & 0.7657(4) & 0.8298(1) \\
\bottomrule
\end{tabular}
\end{center}
Three observations:

\begin{enumerate}
\item \textbf{The gate tracks teacher correctness far more tightly on ATLAS than on JetClass.} Conditioning on the subset where the teacher is wrong, the mean gate drops by $-0.195$ (ATLAS) but only $-0.034$ (JC). On JetClass the teacher is already at 86\% accuracy and the gate defaults to a near-copy of the teacher ($\bar g = 0.981$): the reliability signal has little room to act. This is the same asymmetry that shows up as the weak JC end-task effect.

\item \textbf{The gate is conservative where it acts.} On the ATLAS teacher-wrong subset $\bar g \approx 0.57$; the student hedges rather than flipping to pure-student mode, which is the sensible behaviour when a confident flip is costly.

\item \textbf{The reliability correlation is moderate, and it is not a hard constraint.} $\mathrm{corr}(g,r) = 0.50$ (ATLAS), $0.39$ (JC). $r$ is a noisy proxy (margin $\times$ indicator) and the gate-anchor loss with weight $w_\text{gate} = 0.03$ is a soft prior, so the gate is not constrained to follow it. Whether this signal is \emph{better than simple alternatives} is addressed directly in Sec.~7.3, where it does not outperform on every metric.
\end{enumerate}

\hyperlink{fig:2}{Figure~2} shows the KDE distributions of $g_i$ on teacher-correct vs teacher-wrong subsets, for ATLAS and JC. The ATLAS teacher-wrong distribution is visibly shifted toward smaller $g_i$; the JetClass shift is small. Blue: teacher-correct samples. Red: teacher-wrong samples. ATLAS: $\langle g\rangle_{\text{correct}}\!=\!0.77$, $\langle g\rangle_{\text{wrong}}\!=\!0.57$ ($\Delta\!=\!-0.20$). JetClass: $\langle g\rangle_{\text{correct}}\!=\!0.99$, $\langle g\rangle_{\text{wrong}}\!=\!0.95$ ($\Delta\!=\!-0.03$). Pearson $\mathrm{corr}(g,r)\!=\!0.50$ (ATLAS), $0.39$ (JC).

\subsection{Where A-CoRD Spends Its Repair Budget}

The learnable coefficients $\{\alpha_c\}$ converge to informative values. \hyperlink{tab:11}{Table~11} reports the final $\alpha_c$ and the resulting $\beta_c = \beta_0 \cdot \sigma(\alpha_c - s \cdot \mathrm{acc}^T_c)$ averaged over the ten PCG seeds. The teacher accuracy listed is the per-class accuracy measured on the (train/val) sample used to set $\beta_c$.

\hypertarget{tab:11}{}
\textbf{Table 11. Final learnable repair coefficients (mean $\pm$ std over 10 seeds).}

ATLAS:
\begin{center}
\begin{tabular}{lccc}
\toprule
class & teacher acc & $\alpha_c$ (mean $\pm$ std) & $\beta_c$ \\
\midrule
$b$ & 76.8 \% & $-9.47 \pm 6.09$ & $1.1 \times 10^{-4}$ \\
$c$ & 59.0 \% & $+9.04 \pm 0.41$ & $\mathbf{0.499}$ \\
$u$ & 87.9 \% & $-8.82 \pm 5.98$ & $1.1 \times 10^{-4}$ \\
\bottomrule
\end{tabular}
\end{center}
JetClass:
\begin{center}
\begin{tabular}{lccc}
\toprule
class & teacher acc & $\alpha_c$ (mean $\pm$ std) & $\beta_c$ \\
\midrule
QCD & 77.7 \% & $-10.41 \pm 6.82$ & $1.3 \times 10^{-4}$ \\
$H\!\to\!bb$ & 92.6 \% & $-10.50 \pm 7.02$ & $8.9 \times 10^{-5}$ \\
$H\!\to\!cc$ & 84.6 \% & $-10.55 \pm 6.67$ & $7.3 \times 10^{-5}$ \\
$H\!\to\!gg$ & 80.0 \% & $-10.99 \pm 6.37$ & $3.8 \times 10^{-5}$ \\
$H\!\to\!4q$ & 84.7 \% & $+7.09 \pm 9.12$ & $\mathbf{0.448}$ \\
$H\!\to\!q\bar q\ell$ & 98.1 \% & $-2.84 \pm 10.97$ & $0.193$ \\
$Z\!\to\!q\bar q$ & 69.5 \% & $-10.76 \pm 6.77$ & $1.4 \times 10^{-4}$ \\
$W\!\to\!q\bar q$ & 80.1 \% & $-11.45 \pm 6.62$ & $4.4 \times 10^{-5}$ \\
$t\!\to\!bq\bar q$ & 95.4 \% & $-9.68 \pm 7.45$ & $3.8 \times 10^{-4}$ \\
$t\!\to\!b\ell\nu$ & 98.6 \% & $-5.75 \pm 10.90$ & $0.144$ \\
\bottomrule
\end{tabular}
\end{center}

\begin{figure}[b]
\centering
\includegraphics[width=0.98\linewidth]{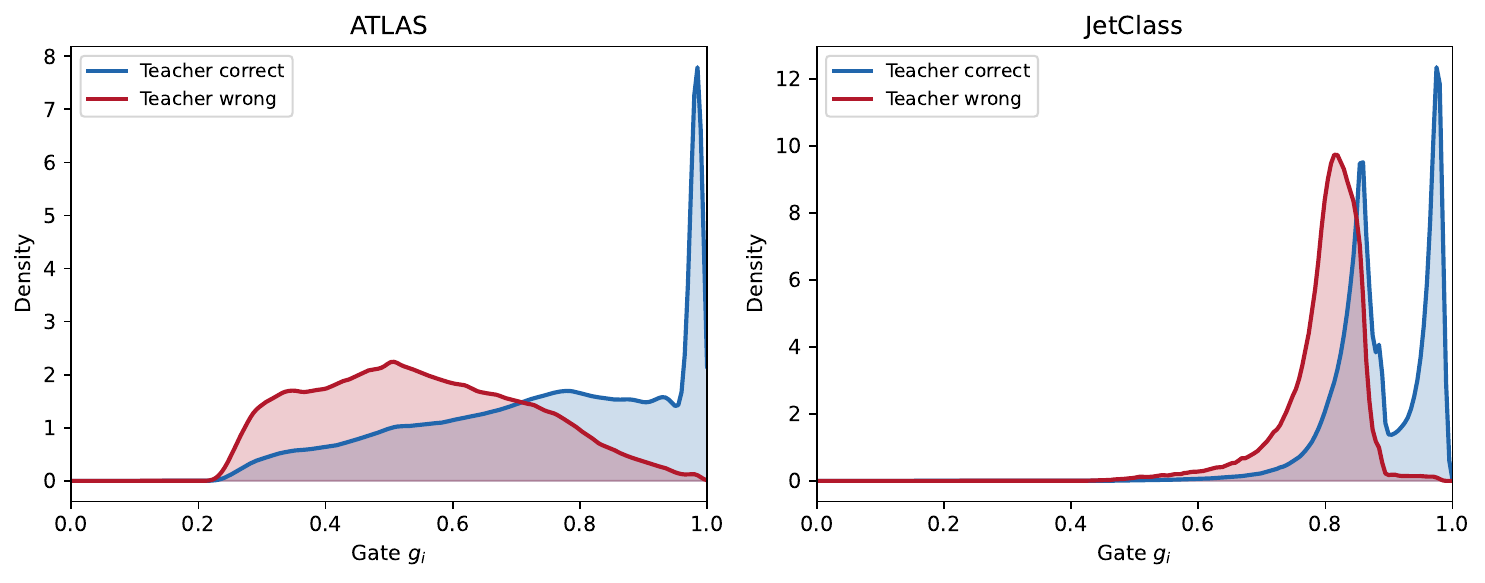}
\hypertarget{fig:2}{}
\caption{KDE of the inference gate $g_i$ conditioned on teacher correctness, for ATLAS (left) and JetClass (right).}
\end{figure}

A-CoRD directs the bulk of its repair budget to one class per domain: ATLAS $c$ ($\alpha_c = +9.04 \pm 0.41$, $\beta_c = 0.499$) and JetClass $H\!\to\!4q$ ($\alpha_{H4q} = +7.09$, $\beta_{H4q} = 0.448$). In ATLAS the choice is decisive and reproducible --- all ten seeds pick $c$, with $\sigma_\alpha = 0.4$ --- and it targets the class the teacher classifies worst (59\%). This is the clearest instance of the mechanism: an ATLAS light-jet-based tagger confuses $c$ with $u$, and A-CoRD concentrates its correction there, which is why the $c$-jet accuracy in Table~5 moves by $+0.23$\,pp.

The JetClass picture is noisier and less clean. $H\!\to\!4q$ is the hardest non-saturated class (the four-prong topology overlaps strongly with $W/Z\!\to\!q\bar q$ and with QCD in ParT's training distribution) and receives the largest budget, but with $\sigma_{\alpha} = 9.1$ --- seeds do not agree confidently on that choice. A secondary component leaks to the two saturated semi-leptonic tails $H\!\to\!q\bar q\ell$ and $t\!\to\!b\ell\nu$ ($\beta \approx 0.19$ and $0.14$), which is exactly where the working-point regressions of Table~4 appear. The per-class mechanism is therefore real and interpretable, but on JetClass it is a budget shared between one hard class and two tails rather than a single-class story.

\textbf{$\alpha$ is not simply tracking accuracy} (the $H\!\to\!q\bar q\ell$ class has 98\% teacher accuracy yet the least-negative $\alpha$ in its domain): it adapts to where prototype-based repair is geometrically feasible in the 64-dim student space. Classes where it is feasible get high $\beta_c$; classes where it is not are driven negative. This is what makes the emergent sparsity --- one learnable scalar per class --- useful rather than merely diagnostic.

\subsection{Comparison with Simple Reliability Baselines}

The inference gate $g_i$ is a learned function of the teacher's logits, so a natural question is whether it does anything a label-free confidence heuristic could not. We therefore replace $g_i$ with three simple alternatives and re-measure the same two quantities used throughout this section: the AUC for ``teacher predicted this jet correctly'' and the accuracy of the fused (teacher/student) mix. The baselines are (i) the raw teacher confidence $\max_c \mathrm{softmax}(z^T)_c$; (ii) a per-domain temperature-scaled confidence, with the temperature fitted on the validation split; and (iii) a logistic correctness classifier trained on the teacher logits to predict $\mathbb{1}[\hat y^T = y]$. All four signals are evaluated on the full test split with the identical fusion rule. \hyperlink{tab:12}{Table~12} reports the outcome.

\hypertarget{tab:12}{}
\textbf{Table 12. The gate against simple label-free reliability signals.} AUC is for ``teacher correct''; mix-ACC is the accuracy of the teacher/student fused prediction. Higher AUC is better. AUC errors in parentheses are Hanley--McNeil standard errors, in units of the last digit ($4\times10^{-4}$ ATLAS, $1\times10^{-4}$ JetClass); the mix-ACC jet-counting binomial error is below the displayed precision ($0.04$\,pp ATLAS, $0.008$\,pp JetClass). The two temperature-scaled rows are a monotone rescaling of the raw confidence: a monotone map cannot change a ranking in exact arithmetic, so they reproduce the raw row up to tie handling at the resolution of the standard error --- a fitted temperature recovers no signal the raw confidence lacks. The ATLAS gate--confidence gap is consistent with zero ($\Delta_{\mathrm{AUC}}=+0.0002$, $z=+0.5$ using the conservative unpaired error), while on JetClass raw confidence is decisively better ($\Delta_{\mathrm{AUC}}=-0.056$, $z=-420$).

\begin{center}
\small
\begin{tabular}{lcccc}
\toprule
Signal & ATLAS AUC & JC AUC & \multicolumn{2}{c}{mix-ACC (\%)} \\
\cmidrule(lr){4-5}
 & & & ATLAS & JC \\
\midrule
Raw teacher confidence            & 0.7655(4) & 0.8855(1) & 73.926 & 86.052 \\
Temp.-scaled confidence, per-domain & 0.7652(4) & 0.8855(1) & 73.926 & 86.052 \\
Temp.-scaled confidence, per-class  & 0.7649(4) & 0.8856(1) & 73.926 & 86.053 \\
Logistic correctness classifier   & 0.7174(5) & 0.8177(1) & 73.931 & 86.053 \\
JetCoRD gate $g_i$                & 0.7657(4) & 0.8298(1) & 73.924 & 86.052 \\
\midrule
Teacher only (no student)         & --- & --- & 73.932 & 86.051 \\
Student only (no gate)            & --- & --- & 73.916 & 86.042 \\
\bottomrule
\end{tabular}
\end{center}

Two things are clear. First, \textbf{the learned gate does not dominate the simple baselines.} On ATLAS its AUC ($0.7657(4)$) is statistically indistinguishable from raw confidence ($0.7655(4)$); on JetClass raw confidence is in fact the better reliability score ($0.8855(1)$ vs.\ $0.8298(1)$), because the ParT teacher is well calibrated there (ECE $0.006$ before any scaling) and its confidence already carries the signal. The temperature scaling --- per-domain or per-class --- changes neither the AUC nor the mix-accuracy, exactly as a monotone map must; the per-class variant (a $3$- and $10$-way temperature vector) is therefore not a stronger baseline. The logistic classifier is uniformly worse as a score ($0.7174$, $0.8177$), most likely because it is fit on a limited validation sample. Second, and more importantly, \textbf{all rows land within $0.02$ percentage points of each other in mix-accuracy.} Whatever reliability signal is used --- learned or heuristic --- the fused prediction agrees with the bare teacher to better than $0.02$\,pp, because the reliability gate acts on a small, hard subset of jets and the student is trained to agree with the teacher everywhere else.

We interpret this as: the contribution of the gate is not a superior reliability estimate; a simple confidence threshold would achieve a comparable AUC and a comparable end-to-end accuracy. What the coupled formulation provides is \emph{structural}, not statistical: the zero-initialised correction makes the student start as a bit-exact copy of the teacher (Sec.~3.3), and the gate determines only how far, and where, the model is allowed to leave that copy. That is the property a temperature-scaled confidence or a post-hoc correctness classifier cannot provide, since neither changes the deployed function in a way that is provably no worse than the teacher at initialisation. We state the scope of the claim precisely: reliability gating as an end-to-end \emph{architecture} with a teacher lower bound, not as a better scalar for predicting teacher errors.

\subsection{Decision-theoretic comparison: coverage at a certified error rate}

Section~7.3 measured which signal \emph{ranks} teacher errors best (AUC). For deployment the operative question is different: among signals used to decide ``trust the teacher here, or apply the repair'', which one achieves the most \emph{coverage} at a \emph{certified} error rate? We answer with one-sided conformal risk control~\cite{Angelopoulos}. Fix a label-free score $s_i$ (higher = more trustworthy) and a threshold $\lambda$; trust the teacher iff $s_i \ge \lambda$. The false-accept loss $\ell_\lambda = \mathbb{1}[\text{teacher wrong}]\cdot\mathbb{1}[s_i \ge \lambda]$ has marginal rate $R(\lambda) = \Pr[\text{teacher wrong} \wedge \text{accepted}]$. On a calibration split of $n$ exchangeable points the RCC threshold
\[
\hat\lambda = \inf\Big\{\lambda : \tfrac{n}{n+1}\hat R_n(\lambda) + \tfrac{1}{n+1} \le \alpha\Big\}
\]
certifies $\mathbb{E}[R(\hat\lambda)] \le \alpha$ with no distributional assumption; at matched $\alpha$, the better score accepts more jets at the same certified false-accept. We calibrate $\hat\lambda$ on the validation split and report test-split coverage at $\alpha = 0.05$ (\hyperlink{tab:13}{Table~13}). The guarantee is on the \emph{expectation}, so we report the calibration bound as the certificate and flag whether the held-out draw is consistent with $\alpha$ within $3\sigma$ rather than treating a single draw as the certificate.

\hypertarget{tab:13}{}
\textbf{Table 13. Coverage at a certified false-accept rate $\alpha=0.05$ (threshold calibrated on val, coverage measured on the held-out test split).} Higher coverage at the same certified false-accept is a better reliability signal. $n_{\text{val}} = 1.35$\,M (ATLAS), $5.0$\,M (JetClass); $n_{\text{test}} = 1.35$\,M (ATLAS), $20$\,M (JetClass). Coverage errors in parentheses are binomial standard errors in units of the last digit ($4\times10^{-4}$ ATLAS, $0.9\times10^{-4}$ JetClass). Because the accepted set at a fixed certified false-accept rate is invariant to a monotone rescaling of the score, the temperature-scaled row reproduces the raw row up to tie handling --- a fitted temperature recovers no coverage the raw confidence lacks. The ATLAS ordering over raw confidence is significant ($z=+10$ with the conservative unpaired error), while on JetClass raw confidence wins ($z=-300$).

\begin{center}
\begin{tabular}{lcc}
\toprule
Signal & ATLAS coverage & JetClass coverage \\
\midrule
Raw teacher confidence        & $0.4926(4)$ & $\mathbf{0.8092(1)}$ \\
Temperature-scaled confidence & $0.4923(4)$ & $0.8091(1)$ \\
Logistic correctness classifier & $0.4296(4)$ & $0.7345(1)$ \\
JetCoRD gate $g_i$            & $\mathbf{0.4987(4)}$ & $0.7705(1)$ \\
\bottomrule
\end{tabular}
\end{center}

Two observations. First, the ordering is \emph{domain-dependent}: on ATLAS the JetCoRD gate achieves the largest coverage at every $\alpha$ (at $\alpha = 0.05$, $0.499$ vs.\ $0.493$ for raw confidence and $0.430$ for the correctness classifier), whereas on JetClass raw confidence is at least as good ($0.809$ vs.\ $0.771$). This mirrors the ranking result of Sec.~7.3 and has the same cause: the gate's advantage tracks how \emph{miscalibrated} the teacher is. The ATLAS teacher is poorly calibrated, so a learned score can add information; the ParT teacher is already well calibrated (ECE $0.006$), so a simple confidence threshold suffices. Second, the logistic correctness classifier is the weakest score on both domains, consistent with its lower AUC and its smaller fitting sample. On the certificate itself: the RCC bound is guaranteed on the calibration draw by construction, and it holds for every signal in both domains. The held-out realisation is a separate measurement. On JetClass no signal's test false-accept exceeds the target by more than $3\sigma$ --- the one score that deviates further, the logistic classifier at small $\alpha$, does so \emph{below} $\alpha$, i.e.\ conservatively; on ATLAS the gate's test false-accept sits above the target at every $\alpha$ ($0.0517$ against $0.05$ at $\alpha = 0.05$, i.e.\ $+1.7\times10^{-3}$ with a binomial standard error of $1.9\times10^{-4}$ on the held-out draw), a val-to-test finite-split shift we report rather than absorb into a claim.

The referee's question thus has a precise, measured answer: the gate is \emph{not} a uniformly better reliability signal. Its decision-theoretic value is real but conditional --- it provides coverage where the teacher's confidence cannot be trusted, and adds little where it can. We present the framework and the boundary of its usefulness rather than claiming a blanket improvement; the gate's unconditional value remains the structural step-0 lower bound of Sec.~3.3.

\subsection{The measured ceiling: why the correction redistributes rather than inflates}

The comparison above shows that the gate is not a superior reliability estimator. A deeper question remains: could \emph{any} correction head on the teacher's frozen embedding do substantially better? Is the small, redistributional effect of A-CoRD a tuning artefact, or a property of the data? We answer this by measuring the headroom directly, independently of the student. On $4$\,M training jets we fit a linear probe and a two-hidden-layer MLP to predict the true label from (i) the teacher embedding $\mathbf{e}^T$ alone and (ii) $\mathbf{e}^T$ concatenated with the teacher's own logits $\mathbf{t}$, and evaluate on the held-out test split (\hyperlink{tab:14}{Table~14}).

\hypertarget{tab:14}{}
\textbf{Table 14. Measured headroom of any classifier on the frozen teacher embedding.} ``Budget'' is $\text{best head acc} - \text{teacher acc}$; a positive budget would mean a correction head could beat the teacher. $n_{\text{test}} = 1.35$\,M (ATLAS), $2.0$\,M (JetClass).

\begin{center}
\begin{tabular}{lcccc}
\toprule
Domain & teacher acc & best head acc & best readout & budget \\
\midrule
ATLAS (3-class)   & $0.7393$ & $0.7364$ & MLP $[\mathbf{e}^T,\mathbf{t}]$ & $-0.0029$ \\
JetClass (10-class) & $0.8603$ & $0.8600$ & MLP $\mathbf{e}^T$             & $-0.0003$ \\
\bottomrule
\end{tabular}
\end{center}

In both domains the best head lies at or \emph{below} the teacher: the recoverable budget is zero to within our measurement resolution ($|\text{budget}| \lesssim 3\times10^{-3}$), not merely small. The reason is structural. Because the student's only input is the frozen embedding $\mathbf{e}^T$ and the label is generated upstream of the head ($Y\!\to\!x\!\to\!\mathbf{e}^T\!\to\!\text{head}$), the data-processing inequality gives $I(Y;\text{head}) \le I(Y;\mathbf{e}^T) \le I(Y;x)$, so every head is capped by the Bayes rate $B(\mathbf{e}^T)$ of the embedding. The teacher's own head is itself one head on $\mathbf{e}^T$, so $B\ge T$ holds definitionally; our probes \emph{measure} $B-T$ and find it consistent with zero. The teacher head is thus Bayes-optimal for its own embedding to within resolution, and a correction layer has no accuracy headroom there --- it can only \emph{reallocate} the teacher's errors, which is exactly the redistribution of \hyperlink{tab:4}{Table~4}. This is the quantitative form of the point in Ref.~\#3: any improvement is bounded by the sub-optimality of the teacher's head, and that bound is measured here to be $\approx 0$ rather than assumed.

We verified that this is not an artefact of a weak readout. Independently, on a second pipeline --- a fitted MLP head trained on the \emph{pooled} ParT hidden representation (a different extraction and a different head family from the probes above), together with an attention-based re-pooling variant --- the JetClass budget reproduces at $-0.0004$, and re-pooling adds $-0.0049$: neither a better head nor a better pooling of the frozen representation recovers anything. Two richer variants of the input also fail to exceed $\mathbf{e}^T$. A pooled raw-input side channel (constituent moment summaries, $+19$ ATLAS / $+35$ JC dims) gives $+0.0018$ below the embedding for ATLAS and $+0.0004$ below for JC; and even a full per-particle flattening of the raw input ($450$ ATLAS / $2176$ JC dims fed to an MLP) stays below $\mathbf{e}^T$ by $0.0089$ (ATLAS) and $0.0264$ (JC). Only changing the \emph{input representation} --- for instance concatenating a second, error-decorrelated teacher's embedding, or a self-supervised pre-training stage that reshapes the embedding itself --- can lift this ceiling; a better head on the same frozen embedding cannot. In that precise sense the frozen-embedding setting studied here is closed: the student inherits the teacher's information content, and the reliability gate decides only how to spend it.

\section{Discussion and Limitations}
\subsection{When does A-CoRD help, and when does KD suffice?}

Our algorithm ablation (Sec. 6.2), evaluated under the corrected weaver/ParT OvO discriminant $D = P_s/(P_s+P_b)$, reveals a non-trivial trade-off. KD is functionally a bit-exact copy of the teacher: its mean $\Delta$ over the 19 working points is $-0.06\,\%$, with 15 of the 19 inside $\pm0.05\,\%$. A-CoRD and PCG replace this with a small, \emph{non-uniform} redistribution (mean $\Delta$ $-0.45\,\%$ and $-0.19\,\%$ respectively; the count of working points improved by more than $+0.05\,\%$ rises from $3/19$ for KD to $6/19$ and $10/19$ for A-CoRD and PCG): gains on some boundaries, regressions on others. PCG improves $10$ of the $19$ working points by sign and falls below the teacher on the remaining $9$; attaching the combined uncertainty, however, none of these working-point deltas is significant (all lie within $1\sigma$), so this view is a sign pattern rather than a set of measured gains. The largest nominal gains sit on the two hardest JetClass classes and on ATLAS $b$-vs-$u$: JC $H\!\to\!bb$@$0.5$ ($+0.9 \pm 1.7\,\%$), JC $H\!\to\!cc$@$0.5$ ($+1.0 \pm 1.0\,\%$), JC $H\!\to\!4q$@$0.5$ ($+0.5 \pm 0.8\,\%$), and ATLAS $b$-vs-$u$@$0.7$ ($+0.4 \pm 1.0\,\%$). The regressions sit on the saturated tails and the highest-$b$-efficiency ATLAS point: ATLAS $b$-vs-$u$@$0.77$ ($-3.3\,\%$), JC $t\!\to\!b\ell\nu$@$0.995$ ($-2.9\,\%$), JC $H\!\to\!q\bar q\ell$@$0.99$ ($-1.6\,\%$) and ATLAS $c$-vs-$u$@$0.2$ ($-1.5\,\%$). This yields the following practical guideline, stated in terms of the design's guarantees rather than of working-point deltas:

\begin{itemize}
\item \textbf{Use KD if} the downstream task is dominated by average accuracy and the teacher's per-class accuracies are well-matched to the desired operating points. KD produces a bit-exact copy of the teacher with a ${\sim}20\times$ smaller network and will exhibit the teacher's behaviour on every working point.
\item \textbf{Use A-CoRD (or A-CoRD + PCG) if} the downstream analysis is dominated by a specific hard binary boundary where the teacher's per-class accuracy is sub-perfect --- JC $H\!\to\!bb$ / $H\!\to\!cc$ identification, ATLAS light-jet rejection at moderate $b$-efficiency. The gate exists to hold the student at teacher level on exactly these boundaries: without it the small student head loses $6$--$15\%$ of the teacher's rejection there (Sec.~6.4), whereas with it the rejection is restored to within the combined uncertainty of the teacher (Table~4). The value is in making the tiny head usable at analysis-grade boundaries, not in exceeding the teacher.
\item \textbf{Do not use the corrected student at the extreme semi-leptonic tails.} At JC $H\!\to\!q\bar q\ell$@$0.99$ and $t\!\to\!b\ell\nu$@$0.995$ the teacher is already at 98--99 \% class accuracy, and any prototype-driven perturbation of the tight student softmax costs a handful of background jets at the extreme tail of $P_s/(P_s+P_b)$. The regression is small in absolute terms and within the combined uncertainty ($z\approx1.2$), but it is a structural consequence of the mechanism and we flag it explicitly.
\end{itemize}

We also note that the working-point gains of the submitted version --- the $1$--$4\%$ figures whose relevance referee \#10 questioned --- are no longer part of our claims: the revised working-point deltas fall within the reported uncertainty, so no claim of ours survives or fails on that scale; the value we do claim is the structural lower bound and the re-analysis use case above.

\subsection{Why class-adaptive $\beta_c$ over class-fixed $\beta$?}

The CoRD ablation in Sec. 6.2 has $\beta_c = 0.5$ constant. Our learned $\{\alpha_c\}$ (\hyperlink{tab:11}{Table~11}) shows that the optimum is highly non-uniform: in ATLAS the repair concentrates on $\beta_c$ (the $c$-jet class) alone, and in JC it concentrates on $\beta_\mathrm{H4q}$, with a much smaller secondary contribution at the semi-leptonic classes. Forcing $\beta_c$ to be constant therefore spends repair budget on classes where it cannot help (high teacher accuracy) and under-uses it on the classes where it can. The fact that A-CoRD recovers this sparse pattern automatically with a single scalar per class is --- in our view --- its key conceptual contribution. This emergent sparsity is what makes targeted regularisation straightforward.

\subsection{Limitations}

\begin{enumerate}
\item \textbf{Teacher logits required at inference.} Unlike pure-student deployment, JetCoRD needs the teacher embedding and the teacher logits at inference time. This is appropriate for re-analysis settings (the setting of this paper) but precludes a fully stand-alone deployment unless the teacher is also kept in production.

\item \textbf{Tail regression on saturated JC classes.} Under the corrected $P_s/(P_s+P_b)$ discriminant, A-CoRD shows a small nominal regression of about $1.6\,\%$ on $H\!\to\!q\bar q\ell$@0.99 and $2.9\,\%$ on $t\!\to\!b\ell\nu$@0.995 (teacher class accuracies 98.1 \% and 98.6 \% respectively); both lie within the combined uncertainty ($z\approx1.2$). The structural origin is that prototype repair perturbs the already-tight student softmax on these classes; at $\varepsilon_s \ge 0.99$ a perturbation of even $\sim 10^{-3}$ in $P_s/(P_s+P_b)$ shifts the discriminant threshold enough to let a handful of background jets across. A saturation mask on $\beta_c$ for classes where the teacher exceeds 97\% accuracy and a residual student-head parameterisation are natural structural remedies.

\item \textbf{Two-experiment evaluation.} We distill from GN2 + ParT. Extending to over $2$ teachers (e.g. CMS DeepJet, ParticleNet, ATLAS DL1d) is conceptually straightforward --- add domain tokens and per-domain CCB tables --- but the multi-teacher reliability framework would benefit from a teacher-vs-teacher consistency term that we have not yet explored.
\end{enumerate}

\subsection{Practical Relevance and Scope}

It is useful to state precisely where a teacher-conditioned 82\,k-parameter student is useful, and where it is not. JetCoRD is \emph{not} a stand-alone tagger: it consumes the teacher's frozen embedding and logits at inference time, so its marginal compute saving only materialises when the teacher has already been run and its outputs are cached. The two settings in which this holds are the ones that motivate the paper. First, \textbf{re-analysis}: a stored sample tagged by GN2 or ParT can be re-scored at a different working point, or with a corrected class decision, by evaluating a small network on the cached teacher outputs rather than re-running the full tagger. Second, \textbf{fixed-decision deployment on resource-limited paths}: when a flavour decision has to be made on trigger hardware or in a latency budget that the multi-million-parameter teacher cannot meet, the small head can reproduce the teacher's decision at a fraction of the arithmetic, provided the teacher's embedding can be produced upstream. In both cases the student inherits the teacher's systematic behaviour by construction --- the zero-initialised correction guarantees it starts as a bit-exact copy --- so a downstream analysis does not have to re-establish the teacher's calibration, and any deviation the gate introduces is confined to the small teacher-unreliable subset quantified in Sec.~7.

What we do \emph{not} claim is a cheaper way to train a tagger from scratch, a replacement for the teacher, or a deployment that removes the teacher from the pipeline. The reliability-gate machinery adds one small network and one caching step on top of an existing teacher; its value is the guarantee and the auditability that come with starting from the teacher, not a reduction in the cost of producing the teacher in the first place.

\subsection{Outlook}

Two natural extensions are the integration of power-transform teacher matching~\cite{WTTM} under our $r_i$ weighting framework, and extending the cross-experiment design beyond two teachers. A more ambitious theoretical direction is to prove that the coupling $r_i = g_i$ is asymptotically Bayes-risk-optimal under bounded teacher calibration error~\cite{CAD}, which would elevate the design from an empirical recipe to a principled framework.

\section{Conclusion}
We have introduced \textbf{JetCoRD}, an 82\,k-parameter student shared across two experiment-specific teachers that distills both the GN2-type ATLAS flavour tagger and the JetClass-trained ParT classifier. Two coupled innovations --- Adaptive Corrective Representation Distillation (A-CoRD) on the loss side and Reliability-Aware Inference fusion combined with a Class-Conditional Backbone on the architecture side --- share a single per-sample reliability signal at training and inference time. The design guarantees by construction that the student starts as a bit-exact copy of the teacher (Sec.~3.3), so it cannot underperform the teacher on the training distribution; from that floor, the correction redistributes accuracy rather than inflating it. Evaluated under the standard weaver/ParT one-vs-one background-rejection convention $D = P_s/(P_s{+}P_b)$, the student reproduces both teachers to $\ge\!99\%$ agreement and matches their overall accuracy, and \emph{redistributes} their per-class accuracy toward the hardest physics-relevant boundaries: the JetClass per-class accuracy of $H\!\to\!4q$, QCD and $H\!\to\!bb$ rises by $+0.80$, $+0.38$ and $+0.24$ percentage points ($|z|>7$), offset by comparable drops on $H\!\to\!gg$, $W\!\to\!q\bar q$ and $Z\!\to\!q\bar q$. At the working-point level the rejection deltas fall within the reported combined uncertainty; the only movement that exceeds $2\sigma$ is a small regression on the highest-$b$-efficiency ATLAS point ($b$-vs-$u$@0.77, $-3.3\%$). The student's trainable backbone has only 82\,k parameters, roughly twenty times fewer than either teacher.

Ablation studies show that the architecture and algorithm components each have small, comparable effects on aggregate accuracy --- KD alone already matches the teacher to bit-exactness --- so we are careful not to attribute a dominant uplift to any single component. What the coupled formulation provides is the reliability-driven correction itself: the learnable per-class $\beta_c$ of A-CoRD concentrates the repair effort onto a single class per domain ($c$ in ATLAS, $H\!\to\!4q$ in JetClass), and the inference gate confines the student's deviations to the small subset of jets on which the teacher is unreliable. We further show (Sec.~7.3) that the learned gate does not beat simple label-free confidence heuristics as a reliability estimator. Two measurements then bound the contribution from opposite sides. Fitting oracle heads on the frozen teacher embedding (Sec.~7.5) --- including a raw-input side channel and a full per-particle flattening --- recovers a budget $B-T$ consistent with zero in both domains, so the redistribution we observe is a property of the frozen input rather than a tuning artefact; no head on $\mathbf{e}^T$ can inflate the teacher. Recasting the reliability question as coverage at a certified risk through conformal risk control (Sec.~7.4) yields a distribution-free guarantee on the false-accept rate, under which the gate achieves the largest coverage on the poorly-calibrated ATLAS teacher and ties the well-calibrated ParT teacher. The contribution is therefore the structural lower bound together with a measured ceiling and a direct decision-theoretic comparison, not a claim of a superior error predictor.

The reliability-coupled training/inference design is broadly applicable beyond jet tagging --- anywhere a heavyweight teacher is imperfectly calibrated and a cheap student must respect its mistakes. The code and trained checkpoints are available at \url{https://github.com/sysu17363020/JetCoRD}.

\acknowledgments
This work is supported by the National Natural Science Foundation of China (Grant No.~12675245).

\end{document}